\documentclass[twocolumn,amssymb,showpacs, amsmath,nobibnotes,nofootinbib,aps,prb]{revtex4}
\usepackage[dvips]{graphicx}
\usepackage[dvips]{color}
\usepackage{subfigure}
\begin{document}\title{The effect of externally applied pressure on the magnetic behavior of Cu$_2$Te$_2$O$_5$(Br$_x$Cl$_{1-x}$)$_2$}
\author{S.J. Crowe}
\author{M.R. Lees}
\author{D.M$^c$K. Paul}
\affiliation{Department of Physics, University of Warwick, Coventry CV4 7AL, United Kingdom}
\author{R.I. Bewley}
\author{J.W. Taylor}
\affiliation{ISIS Facility, Rutherford Appleton Laboratory, Chilton, Didcot, Oxon OX11 0QX, United Kingdom}
\author{G. McIntyre}
\affiliation{Institut Laue-Langevin, 156X, 38042 Grenoble C\'edex, France}
\author{O. Zaharko}
\affiliation{Laboratory for Neutron Scattering, ETHZ \& PSI, CH-5232 Villigen, Switzerland}
\author{H. Berger}
\affiliation{Institut de Physique de la Mati\`{e}re Complexe, EPFL,CH-1015 Lausanne, Switzerland}

\date{\today}

\begin{abstract} 
The effect of externally applied pressure on the magnetic behavior of Cu$_2$Te$_2$O$_5$(Br$_x$Cl$_{1-x}$)$_2$ with $x=$~0, 0.73 and 1, is investigated by a combination of magnetic susceptibility, neutron diffraction and neutron inelastic scattering measurements. The magnetic transition temperatures of the $x=$~0 and 0.73 compositions are observed to increase linearly with increasing pressure at a rate of 0.23(2) K/kbar and 0.04(1) K/kbar respectively. However, the bromide shows contrasting behavior with a large suppression of the transition temperature under pressure, at a rate of -0.95(9) K/kbar. In neutron inelastic scattering measurements of Cu$_{2}$Te$_{2}$O$_{5}$Br$_{2}$ under pressure only a small change to the ambient pressure magnetic excitations were observed. A peak in the density of states was seen to shift from $\sim$~5 meV in ambient pressure to $\sim$~6 meV under an applied pressure of 11.3 kbar, which was associated with an increase in the overall magnetic coupling strength.
\end{abstract}


\maketitle 
Considerable attention has been paid to frustrated quantum spin systems with reduced dimensionality in recent years due to their fascinating ground states and magnetic behavior~\cite{Diep03}. The spin tetrahedral compounds Cu$_{2}$Te$_{2}$O$_{5}$X$_{2}$ (X=Br,Cl) are recent examples~\cite{johnsson} of systems in which competing intra- and inter-tetrahedral interactions give rise to non-trivial ground state and dynamic magnetic behavior along with the possibility that they lie close to a quantum critical point~\cite{gros}.\par

Early studies of Cu$_{2}$Te$_{2}$O$_{5}$X$_{2}$ revealed typical spin-gapped behavior in the magnetization and low lying singlet excitations in Raman spectroscopy~\cite{lemmens}. Below transition temperatures of $T^{Br}_{N}\sim$ 11.4 K and $T^{Cl}_{N}\sim$ 18.2 K (for X=Br and Cl respectively), the compounds form a complicated incommensurate magnetic structure~\cite{zaharko,zaharko2,zaharko3}. Inelastic neutron scattering~\cite{crowe} measurements revealed further complexity in the dynamic behavior, with dispersive excitations associated with the incommensurate order present in both compounds. Theoretical work has investigated the effect of inter-tetrahedral coupling and the relative strengths of exchange interactions in this system~\cite{valenti,brenig,jensen,kotov,whangbo}. However, no theoretical treatment has satisfactorily explained the experimentally observed behavior, and the true nature of the underlying magnetic interactions remains unclear.\par

External pressure has been used as a tool to tune the underlying interactions in many magnetically ordered systems, and has proved to be an invaluable technique for investigating quantum critical points~\cite{matsumoto,pfleiderer2,pfleiderer}. Previous magnetic susceptibility measurements of polycrystalline Cu$_{2}$Te$_{2}$O$_{5}$Br$_2$ under applied pressure have been reported by Kreitlow et al.~\cite{kreitlow} $T^{Br}_{N}$ is reported to quickly decrease under applied pressure and no longer be observable at 8.2 kbar, indicating that Cu$_{2}$Te$_{2}$O$_{5}$Br$_2$ lies close to a non-magnetically ordered phase. In contrast, the temperature of the maxima in the susceptibility ($T_{max}$), which is associated with the overall magnetic coupling strength, is observed by Kreitlow et al. to increase with increasing applied pressure, by up to 25~\% under 8.2 kbar. A structural analysis of Cu$_{2}$Te$_{2}$O$_{5}$Br$_{2}$ under pressure using angle-dispersive x-ray powder diffraction has also been reported~\cite{wang}, with the atomic positions refined in pressures up to 140 kbar. It is observed that the inter-tetrahedral Br-Br distance decreases with increasing pressure and the Cu-Br-Br path becomes slightly more linear, whilst the Cu-Cu distances increase under pressure. In this paper, the effect of externally applied pressure on the magnetic behavior of Cu$_{2}$Te$_{2}$O$_{5}$(Br$_x$Cl$_{1-x}$)$_{2}$ is investigated through a combination of magnetic susceptibility, neutron diffraction and neutron inelastic scattering measurements under pressure. \par

Susceptibility measurements as a function of temperature were performed with a Quantum Design SQUID magnetometer. Hydrostatic external pressure was applied using an easyLab Technologies Mcell 10 pressure cell, using Sn as an in-situ manometer. Neutron diffraction measurements were performed on a single crystal of Cu$_{2}$Te$_{2}$O$_{5}$Cl$_{2}$ with dimensions 5 mm x 2.5 mm x 2.0 mm, on the D10 diffractometer at the ILL. Hydrostatic pressure was applied to the sample using a CuBe clamp cell with Fluorinert pressure medium, and the in situ pressure was determined using a NaCl manometer~\cite{Decker2,Decker1}. Neutron inelastic scattering measurements of polycrystalline Cu$_{2}$Te$_{2}$O$_{5}$Br$_{2}$ were performed on the direct chopper spectrometer HET at ISIS. A CuBe pressure cell was used and the pressure determined from diffraction measurements of a NaCl manometer on PRISMA at ISIS.


\begin{figure}
\centering
\includegraphics[width = 8.0 cm]{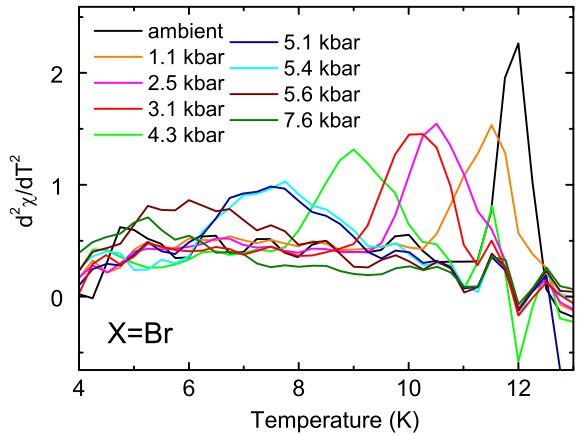}
\caption{The second derivative of the magnetic susceptibility ($d ^2 \chi / d T^2$) as a function of temperature for Cu$_{2}$Te$_{2}$O$_{5}$Br$_2$ under a number of externally applied pressures.}
\label{Brpressderiv}
\end{figure}

\begin{figure}
\centering
\includegraphics[width = 8.0 cm]{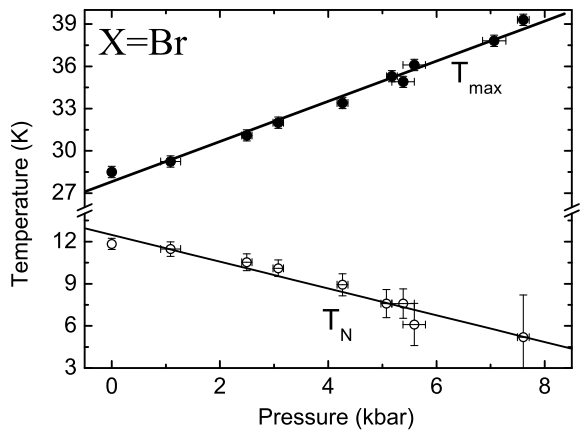}
\caption{The transition temperature, $T^{Br}_{N}$, and the temperature of the maxima in the susceptibility $T_{max}$ of Cu$_{2}$Te$_{2}$O$_{5}$Br$_2$ as a function of pressure.}
\label{Brpress}
\end{figure}
In ambient pressure magnetic susceptibility ($\chi$) measurements of Cu$_2$Te$_2$O$_5$Br$_2$ as a function of temperature, $\chi$ is observed to have a maxima at $T_{max} \sim$~30 K, before dropping abruptly to a much reduced level at lower temperatures~\cite{lemmens}. The transition to long range magnetic order at $T^{Br}_{N}\sim$ 11.4 K corresponds to a step-like feature in $d \chi / d T$, which in turn corresponds to a peak in $d ^2 \chi / d T^2$.\par
Figure~\ref{Brpressderiv} shows $d ^2 \chi / d T^2$ versus temperature for Cu$_{2}$Te$_{2}$O$_{5}$Br$_2$ in an applied magnetic field of 50 kOe as a function of temperature in a number of different applied pressures. Here, the transition temperature (corresponding to the maxima in the peak of $d ^2 \chi / d T^2$) is seen to decrease linearly in applied pressure, from $\sim$~11.5 K in 1.1 kbar, to $\sim$~6.1 K in 5.6 kbar. The peak width increases significantly with increasing applied pressure. By 7.6 kbar there is no longer a clearly defined peak that can be associated with the magnetic transition, although there is a small, broad peak centered at $\sim$~5 K. Unfortunately, below $\sim$~4 K, $d ^2 \chi / d T^2$ becomes rather noisy. This is believed to be because of problems with the background subtraction at low temperatures where the sample susceptibility is very low, and the signal from the pressure cell becomes large due to the presence of paramagnetic impurities in the cell. It is therefore difficult to ascertain whether or not the ordering temperature is suppressed toward $T=$~0 K by further increasing the applied pressure. The pressure dependence of $T^{Br}_{N}$ observed in this data differs somewhat from the results reported by Kreitlow et al.~\cite{kreitlow} In their work, they report that $T^{Br}_{N} \sim$~5 K at a pressure of 3.5 kbar and as a consequence the pressure dependence is rather non-linear. The large number of pressures measured in our work have allowed the observation of a more consistent, linear pressure dependence. In contrast, the temperature at which the maxima in the susceptibility occurs is observed to shift linearly to higher temperatures with increasing pressure, from $\sim$~29 K in 1.1 kbar, to $\sim$~39 K in 7.6 kbar, in good agreement with the work of Kreitlow et al.~\cite{kreitlow} The pressure dependence of $T_{max}$ is displayed in figure~\ref{Brpress}, along with that of $T^{Br}_{N}$. Linear fits give gradients of -0.95(9) K/kbar and 1.42(6) K/kbar for $T^{Br}_{N}$ and $T_{max}$ respectively. \par

\begin{figure}
\centering
\includegraphics[width = 8.0 cm]{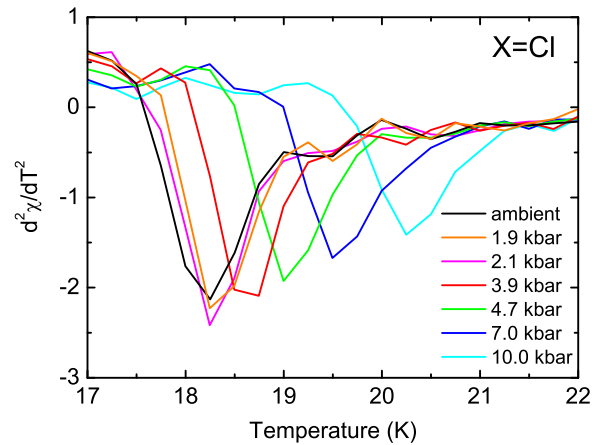}
\caption{The second derivative of the magnetic susceptibility ($d ^2 \chi / d T^2$) as a function of temperature for Cu$_{2}$Te$_{2}$O$_{5}$Cl$_2$ under different externally applied pressures.}
\label{Clpressderiv}
\end{figure}

\begin{figure}
\centering
\includegraphics[width = 8.0 cm]{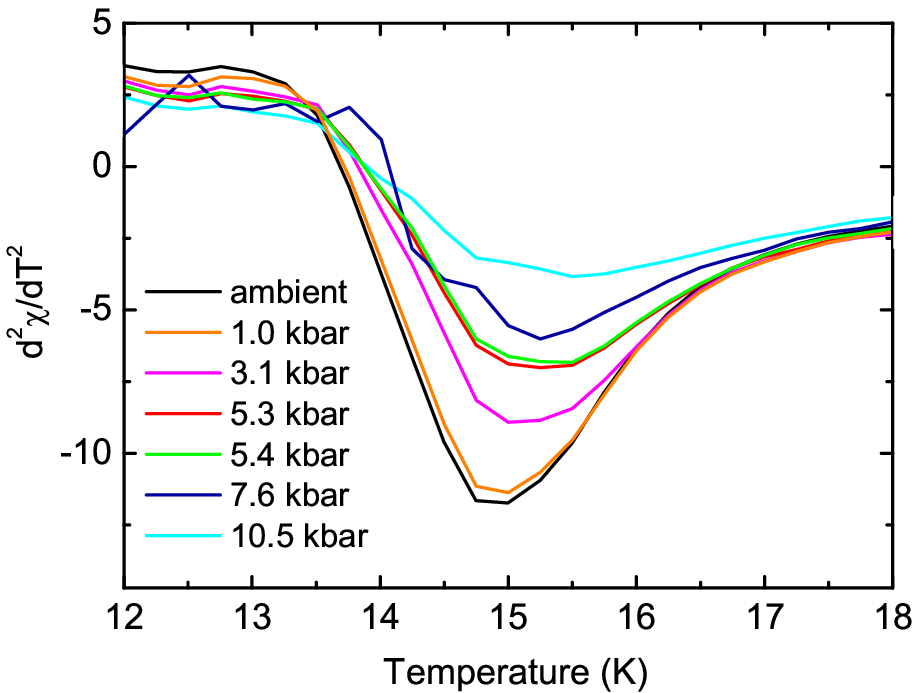}
\caption{The second derivative of the magnetic susceptibility ($d ^2 \chi / d T^2$) as a function of temperature for Cu$_{2}$Te$_{2}$O$_{5}$(Br$_x$Cl$_{1-x}$)$_2$ with $x=$ 0.73, for different externally applied pressures.}
\label{dopepressderiv}
\end{figure}

\begin{figure}
\centering
\includegraphics[width = 8.0 cm]{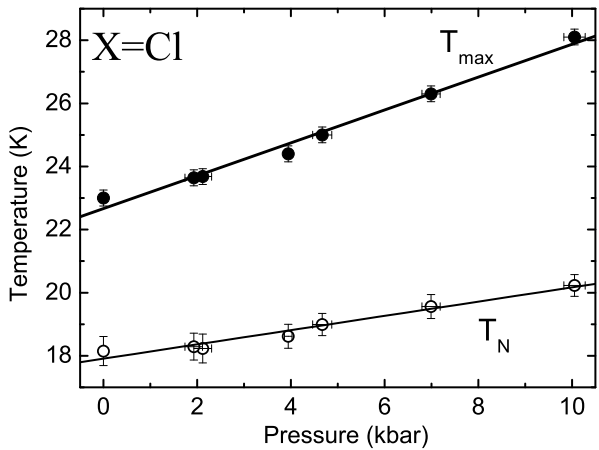}
\caption{The transition temperature, $T^{Cl}_{N}$, and the temperature of the maxima in the susceptibility $T_{max}$ of Cu$_{2}$Te$_{2}$O$_{5}$Cl$_2$ as a function of pressure.}
\label{Clpress}
\end{figure}

Similar measurements have been performed with polycrystalline samples of Cu$_{2}$Te$_{2}$O$_{5}$Cl$_{2}$ and mixed composition Cu$_{2}$Te$_{2}$O$_{5}$(Br$_x$Cl$_{1-x}$)$_2$ with $x=$ 0.73. Figures~\ref{Clpressderiv} and~\ref{dopepressderiv} show $d ^2 \chi / d T^2$ as a function of temperature for a number of different pressures for the chloride and $x=$~0.73 sample respectively. This data was taken in an applied field of 1 kOe for the chloride and 50 kOe for the $x=$~0.73 sample. The transition temperatures ($T^{Cl}_{N}$ and $T^{x=0.73}_{N}$) correspond to the minima in $d ^2 \chi / d T^2$.
In the chloride, $T^{Cl}_{N}$ is observed to increase in a linear fashion with increasing pressure, whilst for the $x=$~0.73 $T^{x=0.73}_{N}$ barely changes with applied pressure, increasing by less than 1 K from ambient pressure to an applied pressure of 10.5 kbar.\par

\begin{figure}
\centering
\includegraphics[width = 8.0 cm]{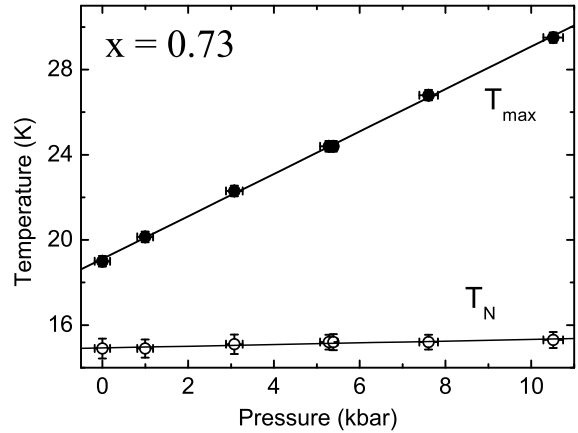}
\caption{The transition temperature ($T^{x=0.73}_{N}$) and the temperature of the maxima in the susceptibility ($T_{max}$) of Cu$_{2}$Te$_{2}$O$_{5}$(Br$_x$Cl$_{1-x}$)$_2$ with $x=$ 0.73 as a function of pressure.}
\label{dopepress}
\end{figure}

The temperature of the maxima in the susceptibility ($T_{max}$) was observed to increase with applied pressure in both samples, from $\sim$~ 23 K in ambient pressure to $\sim$~ 28 K in 10.0 kbar for the chloride and from 19 K in ambient pressure to 30 K in 10.5 kbar for the mixed composition. Figure~\ref{Clpress} plots the pressure dependence of $T^{Cl}_{N}$ and $T_{max}$ for the chloride, with linear fits giving gradients of 0.23(2) K/kbar and 0.52(3) K/kbar respectively. Figure~\ref{dopepress} shows the pressure dependence of $T^{x=0.73}_{N}$ and $T_{max}$ for the $x=$~0.73 compound, both of which show a linear relationship with gradients of 0.04(1) K/kbar and 1.00(1) K/kbar respectively.\par

Neither the transition temperature nor $T_{max}$ respond as strongly to pressure as they do in the case of Cu$_{2}$Te$_{2}$O$_{5}$Br$_{2}$. $T_{max}$ is almost three times as responsive to pressure in the case of X=Br than X=Cl, and $T_{N}$ is almost four times as responsive to pressure for X=Br compared to X=Cl, and, moreover, the pressure dependence acts in the opposite sense for the two compounds. Whilst for the bromide $T_N$ \textit{decreases} with increasing pressure, for the chloride $T_N$ \textit{increases} with increasing pressure, indicating that pressure has a significantly different effect on these two compounds. For the $x=$~0.73 sample, $T_{max}$ behaves under pressure in an intermediate manner to the two end compounds, whilst the small increase in $T^{x=0.73}_{N}$ with pressure is closer to the behavior of the chloride.

\begin{figure}
\centering
\includegraphics[width = 8.0 cm]{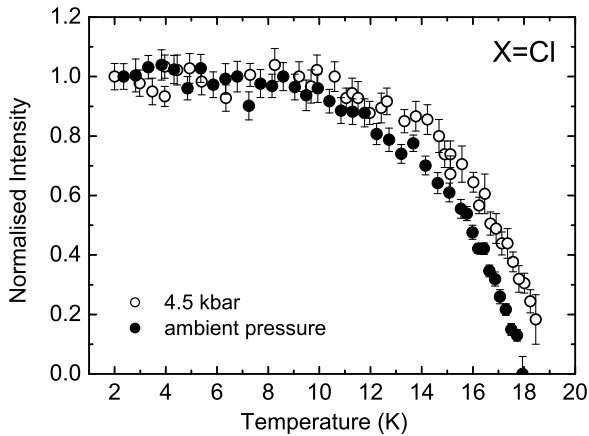}
\caption{Normalized integrated intensity of the $\mathbf{Q}$ = (0.56 0.845 0.5) reflection as a function of temperature for Cu$_{2}$Te$_{2}$O$_{5}$Cl$_{2}$. The filled and empty circles represent the ambient pressure and 4.5 kbar data respectively.}
\label{D10pressure}
\end{figure}

In order to correlate the effect of pressure observed in these macroscopic magnetic susceptibility measurements with a microscopic view of the system, we have also performed neutron diffraction measurements of single crystal Cu$_{2}$Te$_{2}$O$_{5}$Cl$_{2}$ under an applied pressure of 4.5(3) kbar. The modulation vector of the incommensurate magnetic structure ($\mathbf{k'_{Cl}}\approx[-0.15, 0.42, 1/2]$, see ref.~\cite{zaharko}) was not observed to change from that of the ambient pressure structure when under an applied pressure of 4.5 kbar. Figure~\ref{D10pressure} shows the integrated intensity of the $\mathbf{Q}$ = (0.56 0.85 0.5) magnetic reflection over the temperature range 2 - 18.5 K, in which the data taken in ambient pressure (filled circles) and 4.5 kbar (empty circles) have been normalized to the same intensity scale. The data shows a clearly resolved shift in the temperature dependence of the integrated intensity under an applied pressure. The intensity of the magnetic reflection in ambient pressure drops to $\sim$~13 \% of its value at 2 K by 17.75 K, and by 18 K there is no longer a discernable peak. In contrast, in the 4.5 kbar data, the intensity of the magnetic reflection at 18.5 K is still $\sim$~18 \% of its value at 2 K, although at higher temperatures no peak can be resolved above the background level of scattering. The ambient pressure transition temperature of Cu$_{2}$Te$_{2}$O$_{5}$Cl$_{2}$ is $T^{Cl}_N=$ 18.2 K, whilst under a pressure it is observed to shift to $T^{Cl}_N \sim$ 19 K, in good agreement with the susceptibility measurements of Cu$_{2}$Te$_{2}$O$_{5}$Cl$_{2}$ presented above. Similar neutron diffraction measurements of Cu$_{2}$Te$_{2}$O$_{5}$Br$_{2}$ and the $x=$~0.73 composition under pressure have not been possible to date due the difficulty of growing large enough single crystals of these compounds. \par


The effect of applied pressure on the dynamic magnetic behavior of polycrystalline Cu$_{2}$Te$_{2}$O$_{5}$Br$_{2}$ has been investigated using neutron inelastic scattering (NIS) on HET at ISIS. An incident energy of $E_i=$ 18 meV and chopper frequency of 150 Hz were used, giving an accessible $|\mathbf{Q}|$ range of $\sim$~0.5 to 1.5 \AA$^{-1}$ at an energy of 5 meV. Previous NIS measurements~\cite{crowe} of Cu$_{2}$Te$_{2}$O$_{5}$Br$_{2}$ in ambient pressure have revealed the presence of a magnetic excitation with a flat component centered in energy at $\sim$ 5 meV, and a dispersive component centered at $|\mathbf{Q}|\sim$ 0.7 \AA$^{-1}$.

\begin{figure}
\centering
\includegraphics[width = 8.0 cm]{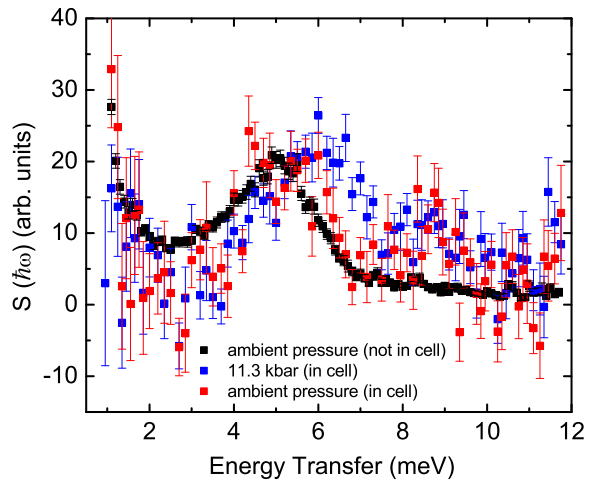}
\caption{$S(|\mathbf{Q}|,\hbar \omega)$ versus energy transfer for Cu$_2$Te$_2$O$_5$Br$_2$, summed over 0.5 \AA$^{-1}<|\mathbf{Q}|<$ 1.5 \AA$^{-1}$. Data for the sample in the pressure cell under 11.3 kbar is shown in blue, for the sample in the pressure cell at ambient pressure in red, and for the sample at ambient pressure and not in the pressure cell in black.}
\label{SEHET}
\end{figure}
\begin{figure}
\centering
\includegraphics[width = 8.0 cm]{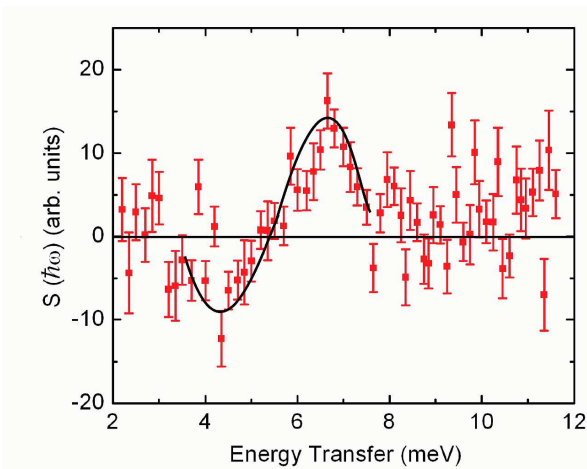}
\caption{$S(|\mathbf{Q}|,\hbar \omega)$ versus energy transfer for Cu$_2$Te$_2$O$_5$Br$_2$, in which the ambient pressure data has been subtracted from the 11.3 kbar data. The data has been summed over 0.5 \AA$^{-1}<|\mathbf{Q}|<$ 1.5 \AA$^{-1}$. The solid line is a guide to the eye.}
\label{SEdiffHET}
\end{figure}

Figure~\ref{SEHET} shows $S(|\mathbf{Q}|,\hbar \omega)$ versus $\hbar \omega$ summed over the $|\mathbf{Q}|$ range 0.5 \AA$^{-1}<|\mathbf{Q}|<$ 1.5 \AA$^{-1}$ for Cu$_{2}$Te$_{2}$O$_{5}$Br$_{2}$ at ambient pressure (red) and under an applied pressure of 11.3 kbar (blue) at T = 4 K. In both cases the background scattering from the pressure cell has been subtracted. The data shown in black in figure~\ref{SEHET} was taken at ambient pressure and outside of the pressure cell, such that the absorption was reduced and better statistics could be obtained in shorter counting times. A peak in the density of states is observed at $\sim$~5 meV in ambient pressure for the case in which the sample is in the pressure cell as well as the case in which it is not in the pressure cell. Indeed, $S(|\mathbf{Q}|,\hbar \omega)$ versus $\hbar \omega$ for both of the ambient pressure measurements are very similar, as expected, although there are small discrepancies, particularly at low energy transfer ($\sim$~1 - 4 meV) $S(|\mathbf{Q}|,\hbar \omega)$, which may be an artifact of the background subtraction due to a possible over-estimation in this region. However, the peak in the density of states for the sample under 11.3 kbar is at $\sim$~6 meV, showing a shift of approximately 1 meV from the ambient pressure data. This shift is more clearly illustrated in figure~\ref{SEdiffHET}, in which the ambient pressure data (taken with the sample in the pressure cell) is subtracted from the 11.3 kbar data.  The experimental setup and instrumental configuration were identical in both cases, hence the subtraction gives the difference in $S(|\mathbf{Q}|,\hbar \omega)$ as a function of energy transfer between measuring under 11.3 kbar and in ambient pressure. The data shows an S-like feature (marked as a solid line in figure~\ref{SEdiffHET} as a guide to the eye), which corresponds to the shifting of the peak under pressure. The negative part of the S-shape feature indicates that there is less intensity in the 11.3 kbar data in the energy region $\sim$~3 - 5.5 meV, in comparison with the ambient pressure data. Similarly, the positive part of the S-shape indicates that there is more intensity in the 11.3 kbar data in the region $\sim$~5.5 - 8 meV compared with the ambient pressure data. From figure~\ref{SEHET}, it appears that there may also be a small broadening of the peak as well as a shift of the center of the peak to higher energy when under applied pressure.

\par


Pressure measurements often provide useful information about the underlying magnetic interactions in a system.
As pressure is applied and the sample volume decreases one would expect, in general, the magnetic coupling strengths to increase due to the closer proximity of the atoms and resultant increased overlap of their orbitals. In particular, for Cu$_{2}$Te$_{2}$O$_{5}$Br$_{2}$ one may naively expect that the effect of pressure would be to push the magnetic behavior toward that of the Cu$_{2}$Te$_{2}$O$_{5}$Cl$_{2}$ compound, which has a 7 \% smaller volume. In this system there are  both intra- and inter-tetrahedral competing interactions, as well as the presence of an antisymmetric DM interaction.  Due to the competition of the different exchange paths, the magnetic ordering is likely to be very sensitive to the relative coupling strengths of the interactions present, which may respond to pressure in different manners. It is therefore important to closely consider the effect of pressure on the structure of the material alongside the effects observed in the magnetic behavior. The structural results of Wang et al.~\cite{wang} for Cu$_{2}$Te$_{2}$O$_{5}$Br$_{2}$ suggest that the inter-tetrahedral exchange interactions $J_a$, $J_b$ and $J_c$ may possibly increase under applied pressure, and the intra-tetrahedral exchange interactions $J_1$ and $J_2$ may possibly decrease under applied pressure (using the exchange interaction notation of Whangbo et al.~\cite{whangbo}). \par

Firstly, consider $T_{max}$, which is observed to increase linearly with increasing applied pressure for each of the samples. In general, $T_{max}$ is associated with the overall coupling strength of the system, which therefore appears to increase under pressure for all compositions. However, it is  not clear which exchange interactions determine the overall magnetic coupling in these materials. In the case of Cu$_{2}$Te$_{2}$O$_{5}$Br$_{2}$, the work of Wang et al.~\cite{wang} suggests that the inter-tetrahedral coupling may increase with applied pressure, whereas the intra-tetrahedral coupling possibly decreases with applied pressure. This indicates that the overall coupling strength may perhaps be determined by the inter-tetrahedral coupling. In the case of Cu$_{2}$Te$_{2}$O$_{5}$Cl$_{2}$ and the doped $x=$~0.73 composition there is no corresponding structural data, and so it is not known how the different exchange paths respond to pressure relative to each other.\par

Now consider the behavior of $T_N$ under pressure for each of the compositions. In contrast to the chloride and $x=$~0.73 sample, the transition temperature of the bromide decreases with increasing pressure, and the compound appears to move toward a non-magnetic phase. However, it is not clear how the pressure dependence of $T^{Br}_N$ develops beyond $\sim$~6 kbar, and whether or not it is completely suppressed at some higher pressure. It has previously been suggested that Cu$_{2}$Te$_{2}$O$_{5}$Br$_{2}$ and Cu$_{2}$Te$_{2}$O$_{5}$Cl$_{2}$ lie in the proximity of a quantum critical point~\cite{lemmens, gros}. These results may therefore suggest that externally applied pressure has the effect of pushing the bromide closer to the quantum critical point, and pushing the chloride away from it. The $x=$~0.73 composition, although somewhat intermediate in its behavior, seems to follow more closely the chloride and is pushed slightly further from the possible quantum critical point with a very small increase in $T^{x=0.73}_N$ under pressure.\par

The contrasting relative behavior of $T_N$ and $T_{max}$ may also give information about the underlying magnetic interactions in the three compounds. Firstly, in a low dimensional system the maxima in the magnetic susceptibility often gives a better indication of the underlying strength of the magnetic interactions than the actual transition temperature, which can reflect weaker 'parasitic' interactions that finally lock in the three dimensional order (examples of this are copper oxy-chlorides CaGdCuO$_3$Cl and Ca$_4$R$_2$Cu$_3$O$_8$Cl$_4$ (R=Gd,Sm)~\cite{sundaresan}). Indeed, for Cu$_{2}$Te$_{2}$O$_{5}$(Br$_x$Cl$_{1-x}$)$_2$ the issue of the relative strengths of the possible interactions, and which determine the low dimensionality or which is dominant in driving the transition to 3D order, has yet to be established.  However, it is clear that in the bromide the interaction driving the 3D order is not the same as that which can be thought of as mediating the overall (possibly low dimensional) coupling because $T_{max}$ and $T^{Br}_N$ act in the opposite sense under pressure. In the Cu$_{2}$Te$_{2}$O$_{5}$Cl$_{2}$ and $x=$ 0.73 samples, $T_{max}$ and $T_N$ both increase under pressure. This may perhaps indicate that the interactions driving the overall magnetic coupling strength also play a role in determining the magnetic ordering temperature. However, it may not be sufficient to think solely in terms of the relative strength of the inter and intra-tetrahedral interactions. Another means by which the magnetic transition in Cu$_{2}$Te$_{2}$O$_{5}$Br$_{2}$ could be suppressed is by an increase in the frustration on the tetrahedra, or possibly a weakening of the DM interaction under pressure, as these are also parameters that may play an important part in the stabilization of a magnetically ordered state in this system. \par

The effect of externally applied pressure on the dynamic magnetic behavior of Cu$_2$Te$_2$O$_5$Br$_2$ appears to be rather more subtle than its effect on the magnetic transition temperature. The most notable change in the dynamics under an applied pressure of 11.3 kbar is a shift of the peak in the density of states of the magnetic excitations to 6 meV, from 5 meV in ambient pressure. \par

In the isolated tetrahedra model, the magnetic excitation present at 5 meV in ambient pressure measurements would correspond to a singlet-triplet spin-gap of $\Delta_{Br}= J_1=$ 5 meV ($J_1 > J_2$). Under pressure, the increase of the peak in the density of states to 6 meV would require the intra-tetrahedral interaction $J_1$ to increase. However, the measurements of Wang et al.~\cite{wang} suggest that under pressure the $J_1$ interaction may perhaps decrease. If, however, the system is considered to consist of isolated square planar units mediated by the inter-tetrahedral $J_a$ and $J_b$ interactions, then the peak energy in the density of states would be determined by the strength of $J_a$ and $J_b$. The work of Wang et al.~\cite{wang} indicates that these inter-tetrahedral $J_a$ and $J_b$ interactions do possibly increase under pressure, which could explain the small shift of the peak in the density of states to 6 meV. However, in a magnetically ordered system there must be some form of coupling between the Cu$^{2+}$ clusters, whether they are tetrahedra or square planar units. Nevertheless, an increase in the overall magnetic coupling with pressure (as observed in the magnetic susceptibility measurements) is consistent with an increase in the energy scale of the magnetic excitations. Unfortunately the statistics of the NIS measurements under pressure are not sufficient to carry out a detailed analysis of the $|\mathbf{Q}|$-dependence in different regions of energy transfer. However, it does appear that the dispersive component of the magnetic excitation in Cu$_2$Te$_2$O$_5$Br$_2$ observed in ambient pressure is still present when a pressure of 11.3 kbar is applied. If the behavior of $T^{Br}_N$ as a function of pressure displayed in figure~\ref{Brpress} is extrapolated to 11.3 kbar, one would expect the transition temperature to be lower than 4 K, and possibly even suppressed close to T = 0 K. Therefore the NIS measurement was most probably performed above the transition temperature, in which case it is proposed that the dispersive excitations are supported by low dimensional order or short range correlations. \par

In conclusion, the effect of externally applied pressure on the magnetic transition temperature ($T_N$) of Cu$_2$Te$_2$O$_5$(Br$_x$Cl$_{1-x}$)$_2$ with $x$ = 0, 0.73 and 1, has been studied using a combination of susceptibility measurements and neutron diffraction. $T_N$ is observed to increase linearly with increasing pressure for the chloride, whilst it decreases rapidly toward T = 0 K in the case of the bromide. For the mixed composition the behavior of $T_N$ is somewhat X=Cl-like, with a very small increase with increasing pressure. In all three compounds the temperature of the maxima in the susceptibility ($T_{max}$), which is associated with the overall magnetic coupling strength, was observed to increase significantly with pressure. Neutron inelastic scattering (NIS) measurements of Cu$_2$Te$_2$O$_5$Br$_2$ revealed a small shift in the peak in the magnetic density of states from $\sim$~5 meV in ambient pressure to $\sim$~6 meV under an applied pressure of 11.3 kbar, which is associated with the increase in the overall magnetic coupling strength. In addition, the dispersive component of the magnetic excitation is still present at 11.3 kbar, and may be supported by low dimensional short range order.

The authors would like to acknowledge the financial support of the UK EPSRC and thank P. Manuel for his help with the PRISMA measurements. We would also like to thank Dr Christophe Thessieu and easyLab for providing the Mcell 10 pressure cell. The sample preparation was supported by the NCCR research pool MaNEP of the Swiss NSF.

\end{document}